\global\def\draftcontrol{0}
 
   \def\versionno{  comments on s-branes   } 
 
\catcode`\@=11 
 
\expandafter\ifx\csname draftcontrol\endcsname\relax\global\def\draftcontrol{0} 
\fi 
 
{\count255=\time\divide\count255 by 60 
\xdef\hourmin{\number\count255} 
\multiply\count255 by-60\advance\count255 by\time 
\xdef\hourmin{\hourmin:\ifnum\count255<10 0\fi\the\count255}} 
\def\draftdate{\number\month/\number\day/\number\year\ \ \ \hourmin } 
  
\newcommand\makepapertitle{\par

  \begingroup 
    \renewcommand\thefootnote{\@fnsymbol\c@footnote}%
    \def\@makefnmark{\rlap{\@textsuperscript{\normalfont\@thefnmark}}}%
    \long\def\@makefntext##1{\parindent 1em\noindent 
            \hb@xt@1.8em{%
                \hss\@textsuperscript{\normalfont\@thefnmark}}##1}%
     \newpage 
     \global\@topnum\z@   
     \@makepapertitle 
     \thispagestyle{empty}\@thanks 
  \endgroup 
  \setcounter{footnote}{0}%
  \global\let\thanks\relax 
  \global\let\makepapertitle\relax 
  \global\let\@makepapertitle\relax 
  \global\let\@thanks\@empty 
  \global\let\@author\@empty 
  \global\let\@date\@empty 
  \global\let\@title\@empty 
  \global\let\title\relax 
  \global\let\author\relax 
  \global\let\date\relax 
  \global\let\and\relax 
  \def\version{\let\version\@version\@gobble} 
} 
\def\@makepapertitle{%
  \newpage 
   \ifnum\draftcontrol=1 {} 
   \version\versionno 
   \vskip 5em%
   \else 
   \hfill\hbox to 3cm {\parbox{4cm}{\@pubnum}\hss}%
   \vskip 5em%
   \fi 
   \begin{center}%
   \let \footnote \thanks 
      {\hskip -0\textwidth \hbox to 1\textwidth%
        {\centerline{\Large\bf{\noindent\@title}}}}%
     \vskip 1.5em%
     {\normalsize
       \lineskip .5em%
       \begin{tabular}[t]{c}%
         \@author 
       \end{tabular}\par}%
     \vskip 1em%
     {\@bstract}%
     \end{center}%
     \vfil 
     \@date%
   \par 
} 
 
\gdef\@pubnum{} 
\def\pubnum#1{%
  \gdef\@pubnum{#1}} 
 
\gdef\@bstract{} 
\def\Abstract#1{%
  \gdef\@bstract{%
   \parbox{\textwidth-0pc}{%
   \centerline{\bf Abstract}\penalty1000 
   \noindent
   \renewcommand\baselinestretch{1.0} 
   {#1}}} 
}

\def\ps@paper{\let\@mkboth\@gobbletwo%
     \ifnum\draftcontrol=1 
        \def\@oddfoot{\hbox to \textwidth{\tiny \versionno \hfil\tiny\draftdate}%
        \hskip -\textwidth \hbox to \textwidth{\hfil\rm\thepage\hfil}}%
     \else\def\@oddfoot{\hbox to \textwidth{\hfil\rm\thepage\hfil}} 
     \fi 
     \let\@evenfoot\@oddfoot 
} 
 
\def\body{\clearpage 
          \pagestyle{paper} 
        } 
\newenvironment{acknowledgments}{%
\vskip 3.25ex 
\noindent {\bf Acknowledgments} 
} 
 
\def\@version#1{\ifnum\draftcontrol=1 
\typeout{}\typeout{#1}\typeout{} 
\vskip3mm\centerline{\hbox{\fbox{\normalsize{\tt DRAFT -- #1 -- } 
                   {\draftdate}}}}\vskip3mm 
\fi} 
\let\version\@version 
\long\def\eqlabel#1{\ifnum\draftcontrol=1 
                    \tag@false  
                    \tag*{(\theequation) \hbox to -0.2cm{\hspace{0cm}\small{#1}\hss}} 
                    \refstepcounter{equation}  
                    \edef\@currentlabel{\theequation} 
                    \ltx@label{#1}          
                    \else 
                    \label{#1} 
                    \fi 
                    } 
\let\st@bibitem\@bibitem 
\let\st@lbibitem\@lbibitem 
\ifnum\draftcontrol=1 
  \def\@bibitem#1{%
    \st@bibitem{#1}\a@@label{#1}\ignorespaces} 
  \def\@lbibitem[#1]#2{%
    \st@lbibitem[#1]{#2}\a@@label{#2}\ignorespaces} 
  \def\a@@label#1{%
    \gdef\a@lab{\smash{\normalfont\small#1}} 
    \ifvmode 
      \if@inlabel 
        \global\setbox\@labels\hbox{%
          \llap{\a@lab\let\a@lab\relax 
                \kern\@totalleftmargin\kern\marginparsep}%
          \box\@labels}%
      \fi 
    \fi} 
\fi 

\documentclass[12pt,letterpaper]{article} 
 
\usepackage{amsmath,amssymb,array,calc,amsthm,rotating,epsfig} 
\usepackage{cite} 
 
\renewcommand\baselinestretch{1.25} 
\renewcommand\section{\@startsection {section}{1}{\z@}%
                                   {-3.5ex \@plus -1ex \@minus -.2ex}%
                                   {2.3ex \@plus.2ex}%
                                   {\normalfont\large\bfseries}} 
\renewcommand\subsection{\@startsection{subsection}{2}{\z@}%
                                   {-3.25ex\@plus -1ex \@minus -.2ex}%
                                   {1.5ex \@plus .2ex}%
                                   {\normalfont\normalsize\bfseries}} 
\renewcommand\subsubsection{\@startsection{subsubsection}{3}{\z@}%
                                   {-3.25ex\@plus -1ex \@minus -.2ex}%
                                   {1.5ex \@plus .2ex}%
                                   {\normalfont\normalsize\it}} 
\renewcommand\paragraph{\@startsection{paragraph}{4}{\z@}%
                                   {-3.25ex\@plus -1ex \@minus -.2ex}%
                                   {1.5ex \@plus .2ex}%
                                   {\normalfont\normalsize\bf}} 
\renewcommand\subparagraph{\@startsection{subparagraph}{5}{\z@}%
                                   {-1.25ex\@plus -1ex \@minus -.2ex}%
                                   {0ex \@plus .2ex}%
                                   {\normalfont\normalsize\it}} 
 
\def\eg{{\it e.g.}} 
 
\def\revise#1       {\raisebox{-0em}{\rule{3pt}{1em}}%
                     \marginpar{\raisebox{.5em}{\vrule width3pt\ 
                     \vrule width0pt height 0pt depth0.5em 
                     \hbox to 0cm{\hspace{0cm}{%
                     \parbox[t]{4em}{\raggedright\footnotesize{#1}}}\hss}}}}

\def\del          {\partial} 
 
\def\ee           {{\it e}}

\def\sqr#1#2{{\vcenter{\vbox{\hrule height.#2pt   
 \hbox{\vrule width.#2pt height#1pt \kern#1pt 
 \vrule width.#2pt}\hrule height.#2pt}}}}

\def\a{\alpha} 
\def\r{\rho}

\def\la{\lambda} 
\def\SO{{\it SO}} 
\def\ISO{{\it ISO}} 
 
\def\app{a_{||}} 
\def\ape{a_\perp} 
\def\Hpp{H_{||}} 
\def\Hpe{H_\perp} 
 
\def\Ppp{P_{||}} 
\def\Ppe{P_\perp} 
\def\kpp{k_{||}} 
\def\kpe{k_\perp}

\catcode`\@=12 
 
\begin{document} 
 
 
\title{Comments on Supergravity Description of S-branes} 
 
\pubnum{%
NSF-KITP-03-30 \\ 
MCTP-03-19 \\ 
hep-th/0305055} 
\date{May 2003} 
 
\author{Alex Buchel$^{1}$ and Johannes Walcher$^{2}$ \\[0.4cm] 
\it $^{1}$Michigan Center for Theoretical Physics \\ 
\it Randall Laboratory of Physics, The University of Michigan \\ 
\it Ann Arbor, MI 48109-1120, USA\\[0.2cm] 
\it $^{2}$Kavli Institute for Theoretical Physics \\ 
\it University of California \\ 
\it Santa Barbara, CA 93106, USA \\[0.2cm] 
}

\Abstract{ 
This is a note on the coupled supergravity-tachyon matter system, 
which has been earlier proposed as a candidate for the effective 
space-time description of S-branes. In particular, we study an 
ansatz with the maximal $\ISO(p+1)\times\SO(8-p,1)$ symmetry, for 
general brane dimensionality $p$ and homogeneous brane distribution 
in transverse space $\rho_\perp$. A simple application of singularity  
theorems shows that (for $p\le 7$) the most general solution with these 
symmetries is always singular. (This invalidates a recent claim in the 
literature.) We include a few general comments about the possibility 
of describing the decay of unstable D-branes in purely gravitational 
terms. 
} 
 
 
\makepapertitle 
 
\body 
 
\version\versionno 
 
\section{Introduction} 
 
Unstable D-branes and their tachyons are a rich source of interesting 
problems in string theory. While the kinematics of tachyon condensation 
and the relation to D-brane charges is by now fairly well understood,  
the decay of unstable branes as a time-dependent process has attracted  
a considerable amount of attention only recently. Guided by intuition 
from ordinary stable D-branes, one is led to expect that this process  
has both a microscopic (or ``open string'') and a macroscopic (or ``closed  
string'') description, which might in some sense be ``dual'' to each other.  
In fact, it was advocated in \cite{gust} that the process of unstable  
brane creation and decay should be viewed as the direct spacelike  
analog of the familiar timelike branes. A recent selection of literature 
on the subject is [1-24]. 
 
An important part of the problem is the effective spacetime description  
of the decay process. Most of the candidate supergravity solutions for  
S-branes that have been written down so far do not satisfy the basic  
conditions on singularity type and global structure. This might be 
due to the restrictiveness of the ansatz used or to the fact that the  
proper set of relevant degrees of freedom has not been identified. A  
hint that the latter might actually be the case comes from the recent 
studies of open string theory with time-dependent boundary perturbations 
\cite{sen,strominger,gust2,maloney}. These results lead to the reasonable  
question whether the decay of unstable branes does actually admit a  
decoupling limit that would be the necessary requirement for  
simplifications. See the end of this note for some additional comments  
on these issues. 
 
It was proposed in \cite{BLW} that a clue to resolving or understanding 
the singularities in the spacetime description of S-branes might lie 
in including the open string tachyon explicitly in the dynamics, coupling 
the tachyon to (super)gravity via the Dirac-Born-Infeld type of action  
commonly known as tachyon matter. To avoid confusion, we note that this 
approach is {\it not} equivalent to studying unstable D-brane probes in 
the background geometry of \cite{gust,cgg,kmp}. In fact, the main point of 
\cite{BLW} was to view the asymptotic geometry of \cite{gust,cgg,kmp} as 
the result of the full gravitational backreaction of the time-dependent 
decay process of (a large number of) unstable D-branes. Translated to say 
the time-like D-branes in type IIB string theory, the question analogous 
to the one raised  in \cite{BLW} would be how to reconstruct the  
classical $p$-brane solution in supergravity from the corresponding boundary 
state \cite{dev}. 
 
The toy model discussed in \cite{BLW} is 4-dimensional Einstein-Maxwell  
theory coupled to tachyon matter on a distribution of one-dimensional 
defects (``D1-branes''). In this toy model, it was shown that all 
solutions are generically singular. In particular, it was shown 
that including the tachyon matter generically destabilizes the  
horizon, turning it into a spacelike (or null) curvature singularity. 
Moreover, it was shown that including the tachyon matter does not  
remove the time-like singularities found previously. Again, we  
emphasize that these statements were based on an analysis that  
accounted for the full backreaction of the tachyon matter on the  
geometry. The most significant drawback of the approach of \cite{BLW}  
was the requirement of $\ISO(1)\times SO(2,1)$ symmetry. This symmetry,  
which is the maximal possible symmetry that one could expect for an 
S-brane of this type, is expected to be broken in the real situation.  
In particular, including tachyon matter would break this symmetry unless 
the branes are smeared uniformly in the transverse space. How to properly 
reduce this symmetry requirement is an important problem. 

\enlargethispage{1.5em} 

Another question is what happens in $10$ dimensions and for general 
brane dimensionality $p$. It was argued in \cite{BLW} that the qualitative  
behavior should be the same as in the toy model, but no explicit 
computations were performed to substantiate this claim.
Recent results presented by Leblond and Peet in \cite{LP} appear
to show that the higher-dimensional analogues of \cite{BLW} do admit 
completely non-singular solutions. Triggered by these results, our 
initial motivation for the present work was to verify this possibility.
However\footnote{The discrepancy of \cite{LP} seems to have both 
an analytical and a numerical origin. See the appendix and the forthcoming 
publication \cite{forth} for comments.}, we will here show that non-singular 
solutions of this system can in fact be easily excluded in general, thus 
confirming the behavior anticipated in \cite{BLW}. These results follow 
from the application of a simple singularity theorem based on the strong 
energy condition satisfied by the tachyon matter, for $p\genfrac{}{}{0pt}{2}
{\displaystyle<}{(=)} 7$. (The case $p=8$ is special and we are unable to 
exclude non-singular solutions on general grounds.) While at first sight 
these results might seem discouraging, we wish to emphasize that we do 
share the hope that tachyon condensation as a dynamical process will 
ultimately admit a complete and physically reasonable description.

\section{Tachyon matter coupled to supergravity} 
 
\subsection{Action and equations of motion} 
 
In this note, we study a system of supergravity fields $S_{\it bulk}$ 
coupled to a (DBI+WZ)-type Lagrangian $S_{\it brane}$ known as tachyon  
matter. We think of $S_{\it brane}$ as representing the degrees of  
freedom of an unstable D-brane system that backreacts on the geometry. 
In this section, we will write out the equations of motion in 
Einstein frame in a convenient form that will make the appearance 
of singularities most obvious. Completely explicit formulas, as well as  
the translation to the string frame, are relegated to the appendix. The  
full action for the coupled system is (the brane has $p+1$ spatial  
dimensions) 
\begin{equation} 
\begin{split} 
S=&S_{\it bulk}+S_{\it brane}\\ 
=&\frac{1}{16\pi G_{10}}\int d^{10}x\ \sqrt{-g}\, 
\Bigl(R-\frac 12 \bigl(\del\phi\bigr)^2-\frac{e^{a \phi}}{2(p+2)!} 
F^2_{p+2}\Bigr)\\ 
&+\frac {\Lambda }{16\pi G_{10}} \int d^{p+2}x_{||}\ \varrho_{\perp}  
\Bigl( -V(T) \sqrt{-A} e^{-\phi}\Bigr)+ 
\frac{\Lambda}{16\pi G_{10}}\int \varrho_{\perp} f(T)\ dT\wedge C_{p+1}\,, 
\end{split} 
\eqlabel{action} 
\end{equation} 
where $a\equiv (3-p)/2$ and  
\begin{equation} 
A_{\mu\nu}=g_{\mu\nu} e^{\phi/2}+\del_\mu T\del_\nu T \,. 
\eqlabel{a} 
\end{equation} 
Here and below we use the symbol $||$, and Greek indices $\mu,\nu, 
\alpha,\beta,\ldots$ for the directions along the unstable brane  
(including time) and the symbol $\perp$ or Roman indices $i,j,\ldots$  
for the transverse directions\footnote{We are assuming here that our  
spacetime is a direct product.}. Capital Roman indices $M,N,\ldots$  
will denote all directions together. 
 
In \eqref{action}, $\varrho_\perp$ describes the distribution of 
branes in the transverse directions. Let us comment. As written in  
\eqref{action}, $\varrho_\perp$ is a form of degree $8-p$, proportional  
to the (appropriately normalized) volume form of the transverse space, 
\begin{equation} 
\varrho_\perp = \rho_\perp \; d^{8-p} x_\perp \,. 
\eqlabel{rhop} 
\end{equation} 
The density $\rho_\perp$ (transforming with the determinant of the  
Jacobian) can depend on the transverse directions, but it is independent  
of the $x_{||}$. Below, we will make an $\SO(p+1)\times\SO(8-p,1)$  
symmetric ansatz, which requires $\rho_\perp$ to satisfy eq.\ \eqref{rour}. 
\footnote{Since we allow for warping of the transverse directions, the  
physical density of branes (branes per unit volume) is given by  
$\rho_{\it phys}$ in 
\begin{equation*} 
\varrho_\perp = \rho_{\it phys} \; {\it vol}_\perp =  
\rho_{\it phys}\;\sqrt{g_\perp}\; d^{8-p}x_\perp \,, 
\end{equation*} 
where $\sqrt{g_\perp}$ depends on the parallel directions (below, it 
will only depend on time). The equation of motion for $\rho_{\it phys}$ 
is ``free streaming'' in the transverse directions.}  
 
As already reviewed in \cite{BLW}, the couplings $V(T)$ and $f(T)$ 
are not known precisely. In \cite{sen} it was argued  
that $V(T)>0$, and  
\begin{equation} 
V(T)\propto e^{-\a |T|/2},\qquad {\rm as}\qquad |T|\to \infty\,, 
\eqlabel{senT} 
\end{equation} 
with $\a=\sqrt{2}$ for superstrings. For numerical analysis of the tachyon  
condensation a convenient choice is $V(T)=1/\cosh (T/\sqrt{2})$, as was  
used in \cite{BLW}, and as recently derived in \cite{kut}. It was noted 
in \cite{BLW} that the singularity argument for the S0-brane was robust as  
to the precise choice of $V(T), f(T)$. Here, we will again find that the 
precise expressions for $V(T), f(T)$ are not important. What matters  
is that $V(T)>0$ and vanishes only as $|T|\to \infty$.  
 
The dilaton, flux, and tachyon equations of motion derived from  
\eqref{action} read, respectively, 
\begin{equation} 
\begin{split} 
0&=\nabla^2\phi+\frac{\Lambda\rho_{\perp}e^{-\phi}V\sqrt{-A}}{\sqrt{-g}} 
\left(1-\frac 14 \left(A^{-1}\right)^{\mu\nu}g_{\mu\nu} e^{\phi/2}\right) 
-\frac{a}{2(p+2)!}e^{a\phi} F^2_{p+2}\\ 
0&=\frac 1{(p+1)!}\epsilon_{\mu\la_2\cdots\la_{p+2}}\ \del_\nu\left( 
\sqrt{-g} e^{a\phi} F^{\nu\la_2\cdots \la_{p+2}}\right)+\Lambda\r_\perp 
f \del_\mu T\\ 
0&=\epsilon_{\mu\lambda_2\cdots\lambda_{p+2}}\!\left[\del_\nu\left(\Lambda 
\r_{\perp} V e^{-\phi}\sqrt{-A}\left(A^{-1}\right)^{\nu\kappa} 
\del_\kappa T\right)-\Lambda\r_{\perp}\sqrt{-A} e^{-\phi}\ \frac{dV}{dT} 
\right]\!-\!\Lambda \r_\perp f F_{\mu\la_2\cdots \la_{p+2}}  
\end{split} 
\eqlabel{eome} 
\end{equation} 
Additionally, we have the Einstein equations 
\begin{equation} 
R_{MN}=T^{(1)}_{MN}+T^{(p+2)}_{MN}+T_{MN}^{brane} \,, 
\eqlabel{eineq} 
\end{equation} 
where $T_{MN}$ denotes the trace reversed energy-momentum tensor of the  
dilaton, form field and tachyon matter, respectively. Explicitly, 
\begin{equation} 
\begin{split} 
T^{(1)}_{MN}&=\frac 12 \del_M\phi\del_N\phi\\ 
T^{(p+2)}_{MN}&=\frac{e^{a\phi}}{2(p+1)!} 
\left(F_{M\cdots} F_N\ ^{\cdots}-\frac{(p+1)}{8(p+2)} 
F_{p+2}^2 g_{MN}\right)\\ 
T^{brane}_{\mu\nu}&=\frac{\Lambda \r_{\perp}V e^{-\phi/2}\sqrt{-A}} 
{16\sqrt{-g}}\left(\left(A^{-1}\right)^{\a\beta}g_{\a\beta}g_{\mu\nu} 
-8\left(A^{-1}\right)^{\a\beta}g_{\a\mu}g_{\beta\nu}\right)\\ 
T^{brane}_{ij}&=\frac{\Lambda \r_{\perp}V e^{-\phi/2}\sqrt{-A}} 
{16\sqrt{-g}}\left(\left(A^{-1}\right)^{\a\beta}g_{\a\beta}g_{ij}\right)\,, 
\end{split} 
\eqlabel{stress} 
\end{equation} 
where dots in the flux stress tensor denote contraction of indices. 
 
\subsection{A homogeneous cosmology} 
 
As in \cite{BLW}, we will impose the maximal possible symmetry on 
our system. As explained in \cite{gust} (see also \cite{cgg,kmp,BLW}) 
this maximal symmetry is $\ISO(p+1)\times \SO(8-p,1)$. Thus, the 
parallel space is flat $(p+1)$-dimensional Euclidean space, and  
the transverse space is the hyperbolic space $H_{8-p}$. But to make 
contact with \cite{cgg,kmp}, and also \cite{LP}, we will write  
out the equations in a slightly more general form, allowing arbitrary  
constant curvature $\kpp$ and $\kpe$ in both the parallel and transverse  
directions. The metric ansatz can then be written in the form 
\begin{equation} 
ds^2 = -dt^2 + a_{||}(t)^2 dx_{||}^2 + a_\perp(t)^2 dx_\perp^2 \,. 
\eqlabel{ansatz} 
\end{equation} 
The ansatz for the flux, dilaton, and tachyon is quite simply 
\begin{equation} 
F_{t,x_1,\cdots,x_{p+1}}=A(t)\ \app^{p+1}\,,\qquad 
\phi\equiv \phi(t)\,,\qquad T\equiv T(t) \,. 
\eqlabel{ans} 
\end{equation} 
We also need to supply the distribution of branes in the transverse 
space. As already indicated, only a homogeneous distribution is consistent 
with the symmetries and simple enough to allow an analytic treatment. For  
instance, if transverse space is $H_{8-p}$, we set 
\begin{equation} 
\r_\perp=\rho_0\sqrt{g_{H_{8-p}}} \,, 
\eqlabel{rour} 
\end{equation} 
where $g_{H_{8-p}}$ is the determinant of $dH_{8-p}^2$, and $\rho_0$ is 
a constant. Despite this high amount of symmetry, we expect that some of  
our statements are actually more general. 
 
The ansatz \eqref{ansatz} is simply a ten-dimensional homogeneous, but 
non-isotropic FRW cosmology. As is well-known, this type of cosmology is  
not generically non-singular, and has at least either a Big Bang or a Big  
Crunch singularity. This knowledge is backed by powerful singularity 
theorems, see, \eg, \cite{hael}. Therefore, if the goal is to construct  
non-singular solutions, one has to make sure that one uses exotic forms  
of matter that allow evading the singularity theorems. We will now show  
that the tachyon matter is, in fact, not quit exotic enough. 
 
Our action \eqref{action} contains three types of matter fields, 
the tachyon, dilaton and RR-form field. The explicit expressions 
for their energy density, $\rho$, and parallel and transverse pressure, 
$\Ppp$ and $\Ppe$, are given in table \ref{matter}. Let us also introduce  
the parallel and transverse Hubble parameters, $\Hpp=\dot\app/\app$ and  
$\Hpe=\dot\ape/\ape$. With this notation, the equations of motion  
for the ansatz \eqref{ansatz} are 
\begin{equation} 
\begin{split} 
-(p+1)\frac{\ddot\app}{\app} - (8-p) \frac{\ddot\ape}{\ape} 
&= \frac 18\bigl(7\rho + (p+1) \Ppp + (8-p) \Ppe\bigr) \\ 
\frac{\ddot\app}{\app} + p \Bigl(\Hpp^2+\frac{\kpp}{\app^2}\Bigr) +  
(8-p)\Hpe\Hpp & = \frac18\Bigl(\rho + (7-p)\Ppp-(8-p)\Ppe\Bigr)\\ 
\frac{\ddot\ape}{\ape} + (7-p)\Bigl(\Hpe^2+\frac{\kpe}{\ape^2}\Bigr)+  
(p+1)\Hpp\Hpe & = \frac18\Bigl(\rho + p\Ppe-(p+1)\Ppp\Bigr)\,. 
\end{split} 
\eqlabel{FRW} 
\end{equation} 
 
\begin{table}[t] 
\begin{center} 
\begin{tabular}{|l|c|c|c|} 
\hline  
& $\rho$ & $\Ppp$ & $\Ppe$ \\ 
\hline 
Tachyon & $\displaystyle\frac{\lambda V(T) \ee^{\phi(p/4-1/2)}} 
{2\ape^{8-p}\sqrt{\Delta}}$ & $\displaystyle-\rho\Delta= 
-\frac{\lambda V(T)\ee^{\phi(p/4-1/2)}\sqrt{\Delta}}{2\ape^{8-p}}$&0\\ 
Dilaton & $\frac14\dot\phi^2$ & $\frac14\dot\phi^2$ & $\frac14\dot\phi^2$ \\ 
RR form & $\frac14\ee^{a\phi}A^2$ & $-\frac14\ee^{a\phi}A^2$ & 
$\frac14\ee^{a\phi}A^2$\\ 
\hline 
\end{tabular} 
\caption{Energy density and pressure \eqref{stresst} for the three matter 
fields in \eqref{action} in the cosmology \eqref{ansatz}. Here, 
$\Delta = 1 - \ee^{-\phi/2}\dot T^2$, and $\lambda=\r_0\Lambda$.} 
\label{matter} 
\end{center} 
\end{table} 
 
{} As is familiar, the equations \eqref{matter} are not independent. 
The constraint (Friedmann equation) is 
\begin{equation} 
\frac{p(p+1)}2\bigl(\Hpp^2 + \frac{\kpp}{\app^2}\bigr) 
+ \frac{(7-p)(8-p)}2\bigl(\Hpe^2+\frac{\kpe}{\ape^2}\bigr) 
+ (p+1)(8-p) \Hpp\Hpe = \rho 
\eqlabel{friedmann} 
\end{equation} 
and is consistent with the equations of motion \eqref{FRW} precisely 
if the energy is covariantly conserved 
\begin{equation} 
\dot\rho = -(p+1)\Hpp (\rho+\Ppp) - (8-p)\Hpe (\rho+\Ppe) \,. 
\eqlabel{consistency} 
\end{equation} 
To check this, we record here the equations for the energy density and  
pressure that follow from the corresponding equations of motion (eq.\  
\eqref{explicit} in the appendix is helpful for this). We have for the  
tachyon 
\begin{equation} 
\dot\rho^{\rm tach} = -(p+1)\Hpp(\rho^{\rm tach}+\Ppp^{\rm tach})  
-(8-p)\Hpe \rho^{\rm tach}-\frac{\lambda A f}{2 \ape^{8-p}} \dot T  
+ \frac{\dot\phi\rho^{\rm tach}}{4}  
+ (3-p) \frac{\dot\phi\Ppp^{\rm tach}}{4} \,. 
\eqlabel{tachyonrho} 
\end{equation} 
The first two terms are of course nothing but the covariant derivative 
of the energy-momentum tensor, while the other terms describe energy 
exchange with the other matter fields. For the dilaton the corresponding  
equation is 
\begin{equation} 
\dot\rho^{\rm dil} = -\bigl((p+1)\Hpp +(8-p)\Hpe\bigr)(\rho^{\rm dil}+ 
P^{\rm dil})+a\ \rho^{\rm RR} \dot\phi  
-\frac{\dot\phi\rho^{\rm tach}}{4}  
- (3-p) \frac{\dot\phi\Ppp^{\rm tach}}{4} \,, 
\eqlabel{dilatonrho} 
\end{equation} 
while for the form field we have 
\begin{equation} 
\dot\rho^{\rm RR} = -(8-p)\Hpe(\rho^{\rm RR} + \Ppe^{\rm RR}) 
+ \frac{\lambda A f}{2 \ape^{8-p}} \dot T  
- a\ \rho^{\rm RR} \dot\phi \,. 
\eqlabel{RRrho} 
\end{equation} 
The consistency condition \eqref{consistency} follows trivially by taking the 
sum of \eqref{tachyonrho}-\eqref{RRrho}. 
 
\subsection{A singularity theorem ($p<7$)} 
 
We wish to show that solutions of \eqref{FRW} are generically singular. 
For a textbook argument, consider the average warp factor $a$, defined by 
\begin{equation} 
a^9 = \app^{p+1} \ape^{8-p} \,. 
\end{equation} 
We have 
\begin{equation} 
9\frac{\dot a}a= 9H= (p+1) \Hpp+ (8-p)\Hpe \,, 
\eqlabel{hdef} 
\end{equation} 
and using \eqref{FRW}, we find, 
\begin{equation} 
\begin{split} 
9\frac{\ddot a}{a} &= (p+1)\frac{\ddot\app}{\app} + (8-p)\frac{\ddot\ape}\ape 
-\frac{(p+1)(8-p)}9 (\Hpp-\Hpe)^2 \\ 
&= -\frac 18\bigl(7\rho + (p+1) \Ppp + (8-p) \Ppe\bigr) 
-\frac{(p+1)(8-p)}9 (\Hpp-\Hpe)^2 \,, 
\end{split} 
\eqlabel{raychaudhuri} 
\end{equation} 
Now from table \ref{matter}, one easily deduces that the combination  
$7\rho+(p+1)\Ppp+(8-p)\Ppe$ on the right hand side of \eqref{raychaudhuri}  
is always positive for $p<7$. This is a {\it strong energy condition}, and it  
implies that $\ddot a$ will be always be negative. Thus, if we consider a  
contracting phase of the universe with $H<0$, \footnote{The case $H>0$ is 
similar, with a singularity in the past.} the slope of $a$ is also  
negative and cannot increase. This implies that $a$ will reach zero in  
a finite time bounded above by $1/|H|$. Intuitively, this contraction  
of the universe will lead to a diverging energy density, hence one  
expects a curvature singularity. While this is not immediate from what  
we have said so far, and there could still be a continuation beyond the  
point where $a=0$, at the very least, the coordinate system \eqref{ansatz}  
will not cover all of spacetime. 
 
\subsection{More careful arguments $(p<7)$} 
 
First of all let us note that asymptotically flat initial condition as  
$t\to -\infty$ for the S-brane cosmology implies that $H(t)<0$ as  
$t\to -\infty$. Indeed, asymptotic flatness of the $\ISO(p+1)\times 
\SO(8-p,1)$ symmetry preserving ansatz \eqref{ansatz} implies  
\begin{align} 
\app\Big|_{t\to -\infty}&\to {\rm constant}, & 
\dot\app\Big|_{t\to -\infty}&\sim\ o\left(\frac 1t\right)\,,\notag\\ 
\ape\Big|_{t\to -\infty}&\to -t, & 
\dot\ape\Big|_{t\to -\infty}&\to -1\,, 
\eqlabel{incond} 
\end{align} 
thus $H$ as defined by \eqref{hdef} satisfies 
\begin{equation} 
H\Big|_{t\to -\infty}=\frac{8-p}{9 t}+o\left(\frac 1t\right)<0\,. 
\eqlabel{hbound} 
\end{equation} 
Moreover, from \eqref{raychaudhuri}, we deduce 
\begin{equation} 
\dot H + H^2 \le 0 \,, 
\end{equation} 
and as a consequence, we obtain the estimate 
\begin{equation} 
H \le \frac{1}{t-t_s}\,, 
\end{equation} 
as long as $t$ is less that the integration constant $t_s$. 
Thus we conclude that $H$ will diverge at some finite time $t=t_s\,$, 
\begin{equation} 
H\Big|_{t\to t_s}\to -\infty\,. 
\eqlabel{singapp} 
\end{equation} 
The arguments we have given so far do not imply that there is an 
actual physical singularity at $t=t_s$. The simplest way to show that 
this divergence is not just a signal of breakdown of the coordinate 
system \eqref{ansatz} would be to establish that the Ricci scalar 
(or any other scalar curvature invariant) diverges as one approaches 
$t=t_s$. While we  expect this to be the case, we have not been  
able to find a convincing argument. It is  possible 
that while the universe is overall contracting, it is expanding in one 
of the two directions at the approach of $t=t_s$. This expansion could  
dilute the energy density so much as to prevent a curvature singularity. 
One would then hope that this is just a coordinate (\eg, Milne) singularity  
and that the spacetime admits an extension and a continuation beyond  
$t=t_s$. One desirable feature of this would be that the additional parts  
of spacetime are not in the Cauchy development of the $t\to -\infty$ 
asymptotic infinity, as one would expect for an S-brane. But even 
if this were the case, it appears that the timelike geodesics associated  
with energy flow of our matter are necessarily incomplete \cite{hael}. 
Additional input would be needed to make sense of the singularity. 
 
We also note that our assumptions for proving the existence of this  
singularity were rather weak, and that, in particular, we did not  
assume invariance under time reversal $t\leftrightarrow -t$. Since the  
arguments only rely on the strong energy condition, which is satisfied  
for $p<7$, it is  reasonable to expect that similar statements about  
singularities will hold even without assuming any particular symmetry. 
 
\subsection{$p=7$ and $p=8$ cases} 
 
Even though the strong energy condition does not hold for $p=7$, this 
case can be treated in essentially the same way\footnote{Of course 
one should find a (new) relevant ``singularity theorem''.} as in the  
previous subsection. The technical details are discussed in the appendix. 
 
The $p=8$ case is special: this S-brane is realized as the gravitational  
backreaction of the time-dependent decay of the unstable D9-brane. 
Since the D9-brane is space filling, there is no transverse space,  
and thus brane smearing \eqref{rour} is irrelevant. One is then 
dealing with a ten-dimensional homogeneous and isotropic flat FRW  
cosmology. The tachyon matter close to the top of the tachyon  
potential and also the RR-form act effectively like a positive cosmological  
constant with $\rho=-\Ppp=-P>0$. However, since $\kpp=0$, one cannot 
reasonably hope for a deSitter-like bounce. The best one can expect 
is an infinite period of inflation if tachyon matter and RR-form 
dominate over the dilaton for early/late times (this could be viewed as 
a ``half S-brane''). We have investigated this question numerically 
, but have been unable to identify a solution  
of this kind. What seems to happen is that the dilaton is always activated  
so rapidly that its positive pressure induces the collapse of spacetime. 
Analytically, we have been able to exclude bouncing solutions that 
are invariant under time reversal, see eq.\ \eqref{conste} in the  
appendix.

\section{Final Comments} 
 
We have seen that the general time-dependent solution of the coupled 
supergravity-tachyon matter system with maximal symmetries is 
singular. This confirms the expectations that the toy model considered  
in \cite{BLW} in fact did capture most of the essential physics of 
the problem. One might ask the question whether this statement would 
still hold true if one were to relax the requirements of symmetry. 
Here one can imagine breaking the maximal $\ISO(p+1)\times\SO(8-p,1)$  
symmetry either to $\ISO(p+1)\times\SO(8-p)$ (by removing the smearing), 
or even further by breaking the $\ISO(p+1)$ S-brane world volume symmetry, 
for instance by allowing more generic, spatially inhomogeneous  
time-dependent profiles of the tachyon field on the unstable D-brane (see,  
\eg, \cite{lnt} for a worldsheet approach to this situation). It is  
possible that since the arguments of the singularity theorem presented  
here (in the context of the maximally symmetric case) rely on the strong  
energy condition for the supergravity and tachyon matter in the decay of  
the unstable brane, the spatially inhomogeneous decay would still describe  
a singular cosmology in the supergravity approximation. 
 
We conclude this note with a few incomplete comments about the possibility  
of describing the decay process of unstable D-branes in a classical  
supergravity theory. The first observation that it is really the  
supergravity approximation that is at fault in generating the singularities  
comes from the origin of the singularity in the toy model of \cite{BLW}.  
In \cite{BLW} the space-like singularity of the S0-brane was attributed to  
the divergence of the tachyon matter energy density (but {\it not} to the  
infinite  energy\footnote{Note the energy density associated  
with the unstable brane distribution is finite.}  from smearing). Furthermore, 
this energy density diverged precisely because the time derivative of the 
tachyon field was getting very large 
\begin{equation} 
\dot T\propto \frac{1}{\sqrt{t-t_s}}\qquad {\rm  as}\qquad t\to t_s+0 \,, 
\eqlabel{tdot} 
\end{equation}  
that is the tachyon was ``rolling too fast''. But large gradients imply  
that the DBI approximation\footnote{The derivation of tachyon matter  
effective action in \cite{kut} is valid only for slowly varying tachyon  
profiles.}, used to couple open string tachyon to the closed string  
background in \eqref{action}, is invalid at times $t\lesssim t_s$.  
In other words, as the tachyon approaches the top of the potential,  
the higher derivatives of the tachyon field in the effective action  
become more and more important. The latter indicates that massive string  
modes might not decouple.  
 
A possibly related observation was made in \cite{strominger,gust2}, in  
which the quantum open string creation in Sen's rolling tachyon background 
(timelike Liouville theory) was computed, and found to diverge due 
to the exponentially growing density of massive {\it open} string 
states. This shows that the tachyon should roll much faster at initial  
times than deduced from the tachyon matter action. These results also  
seem to point to the incompleteness of the open string description and  
to the importance of closed strings. 
 
The closed string couplings of the rolling tachyon worldsheets were 
studied in \cite{sen,oksu,chen,llm}. Most recently, it was shown 
in \cite{llm} that the total amount of closed string radiation is {\it finite} 
in a bosonic string theory, at least for large $p$. This is intriguing, 
as it suggests that an unstable D-brane does not quite manage to  
completely decay into closed strings at the linearized level. One  
would again be led to the conclusion that some mysterious 
form of tachyon matter must intervene as the final state of tachyon 
condensation. The results of \cite{llm} also indicate that it is 
the massive {\it closed} strings that carry away most of the energy. 
 
The computation of \cite{llm} was done for a single unstable D-brane. 
Restoring the factor of $N$ if there are $N$ unstable branes shows 
that the total amount of closed string radiation behaves as $N^2$, so 
that the ratio of radiated energy to initially present brane tension is 
given by 
\begin{equation} 
\# \frac{N^2}{N/g_s} = \# g_s N \,, 
\eqlabel{maldacena} 
\end{equation} 
where $\#$ is a number of order 1. Thus, if  $g_s N$ is large, 
the radiated energy computed in the linearized approximation will  
always exceed the initially present energy, meaning that the 
backreaction on the tachyon has to be taken into account. Moreover,  
the results of \cite{oksu} show that the coupling to massive closed string  
modes, while exponentially growing at late times, is actually small  
at early times. Therefore, these results do not exclude the possibility  
that there is a limit in which only the first few low-lying closed 
string modes are excited, very early on in the decay process. 
Given \eqref{maldacena}, the limit in question appears to be the 
usual Maldacena-type limit of large $g_s N$. 
 
A very important issue\footnote{We thank Joe Polchinski for useful 
discussions on this. Similar points have been made by Martin Kruczenski.} 
is whether the decay of an unstable D-brane leads to a classical final  
state at all or whether the decay process is inherently quantum  
mechanical, possibly with a thermal final state. For instance, the decay 
of an unstable D-brane might liberate so much energy that a black hole 
is formed before this energy can escape to infinity, in particular 
in the large $g_s N$ limit. The recent computations of \cite{maloney}  
should help settling these issues. 
 
In any case, we currently feel that the most promising avenue for making  
progress on the supergravity description of S-branes (if it exists) is to  
relax the symmetry requirements. It particular, it would be interesting 
to improve/generalize the singularity analysis presented in this note to  
the problem of the spatially inhomogeneous decay of unstable D-branes  
\cite{prog}.

\begin{acknowledgments} 
We would like to thank Gary Horowitz, Finn Larsen, Alex Maloney, 
Joe Polchinski, Arkady Tseytlin, and Henry Tye for useful discussions. We 
are grateful to Frederic Leblond and Amanda Peet for important communications.
J.W.\ would like to thank the Michigan CTP for hospitality during the 
workshop on time-dependent backgrounds, where the initial questions  
addressed in this note were raised. The work of A.B.\ is supported in  
part by the U.S. Department of Energy. The work of J.W.\ is supported in  
part by the NSF under Grant No.\ PHY99-07949. 
\end{acknowledgments} 
 
\begin{appendix} 
 
\section{Appendix} 
 
\subsection{Explicit equations} 
Here we  spell out a few more 
details of the equations of motion \eqref{eome} and \eqref{eineq}. 
In a slightly more general gauge than \eqref{ansatz}, the ansatz reads 
\begin{equation} 
\begin{split} 
&ds_E^2=-c_1(t)^2 dt^2+c_2(t)^2 dx^2_{p+1}+c_3(t)^2 dH_{8-p}^2\\ 
&F_{t,x_1,\cdots,x_{p+1}}=A(t)\ c_1 c_2^{p+1}\\ 
&\phi\equiv \phi(t),\qquad T\equiv T(t) \,. 
\end{split} 
\eqlabel{ansatze} 
\end{equation} 
Again, we introduce 
\begin{equation} 
\Delta=1-e^{-\phi/2}c_1^{-2}\left(T'\right)^2,\qquad \lambda\equiv  
\r_0\Lambda \,, 
\eqlabel{delta} 
\end{equation} 
and explicitly evaluate the equations\footnote{Prime denotes derivative  
with respect to $t$. } 
\begin{equation} 
\begin{split} 
&\left(e^{a\phi} c_3^{8-p} A \right)'=\la f T'\\ 
&\frac{1}{c_1 c_2^{p+1}c_3^{8-p}}\left( 
c_1^{-1}c_2^{p+1}c_3^{8-p}\phi'\right)'=\frac 12 a e^{a\phi} A^2 
+\frac{\la V e^{\phi(p/4-1/2)}}{4 c_3^{8-p}}\left((3-p)\Delta^{1/2} 
-\Delta^{-1/2}\right)\\ 
&0=c_1 c_2^{p+1}\left(\frac{dV}{dT} e^{\phi(p/4-1/2)}\Delta^{1/2}+A f\right) 
+\left(c_1^{-1}c_2^{p+1}V e^{\phi(p/4-1)}\Delta^{-1/2}T'\right)'\,. 
\end{split} 
\eqlabel{explicit} 
\end{equation} 
For the Einstein equations we have the nontrivial Ricci components 
\begin{equation} 
\begin{split} 
c_1^{-2}\ R_{tt}&=-\frac{p+1}{c_1 c_2}\left(\frac{c_2'}{c_1}\right)' 
-\frac{8-p}{c_1 c_3}\left(\frac{c_3'}{c_1}\right)'\\ 
c_2^{-2}\ R_{\mu\mu}&=\frac{1}{c_1 c_2}\left(\frac{c_2'}{c_1}\right)' 
+\frac{p}{c_2^2}\left(\frac{c_2'}{c_1}\right)^2+\frac{8-p}{c_1^2} 
\left(\frac{c_2'}{c_2}\right)\left(\frac{c_3'}{c_3}\right)\\ 
c_3^{-2}\ R_{ii}&=\frac{1}{c_1 c_3}\left(\frac{c_3'}{c_1}\right)' 
+\frac{7-p}{c_3^2}\left(\left(\frac{c_3'}{c_1}\right)^2- 
1\right)+\frac{p+1}{c_1^2} 
\left(\frac{c_2'}{c_2}\right)\left(\frac{c_3'}{c_3}\right)\,, 
\end{split} 
\eqlabel{ricci} 
\end{equation} 
and the stress tensor 
\begin{equation} 
\begin{split} 
c_1^{-2}\ T_{tt}&=\frac 12 \left(\frac{\phi'}{c_1}\right)^2 
+\frac{e^{a\phi}(7-p)}{16} \  A^2+\frac{\la V e^{\phi(p/4-1/2)}} 
{16 c_3^{8-p}}\left(7\Delta^{-1/2}-(p+1)\Delta^{1/2}\right)\\ 
c_2^{-2}\ T_{\mu\mu}&= 
-\frac{e^{a\phi}(7-p)}{16} \  A^2+\frac{\la V e^{\phi(p/4-1/2)}} 
{16 c_3^{8-p}}\left(\Delta^{-1/2}+(p-7)\Delta^{1/2}\right)\\ 
c_3^{-2}\ T_{ii}&= 
\frac{e^{a\phi}(1+p)}{16} \  A^2+\frac{\la V e^{\phi(p/4-1/2)}} 
{16 c_3^{8-p}}\left(\Delta^{-1/2}+(p+1)\Delta^{1/2}\right)\,. 
\end{split} 
\eqlabel{stresst} 
\end{equation} 
The full system of equations is overdetermined (there is a standard 
first order constraint analogous to the  
Friedmann equation \eqref{friedmann}),  
and we have explicitly verified that the complete system is  
consistent.  
 
The ``natural'' S-brane solution in supergravity  
would be invariant under the time-reversal $t \leftrightarrow -t$,  
with the tachyon  sitting at the top of the potential at $t=0$.  
This implies that at $t=0$ we would like to have 
\begin{equation} 
\phi'=A=c_2'=c_3'=0 \,. 
\eqlabel{init} 
\end{equation} 
The same boundary conditions were claimed in \cite{BLW} to be inconsistent  
for describing the S$0$-brane in the toy model. It was also mentioned  
in \cite{BLW} and argued in \cite{BW} that the result should also hold  
with coupled dilaton and general brane dimensionality $p$. Indeed, here  
we find that the constraint equation evaluated with \eqref{init} implies 
\begin{equation} 
\Bigl[\la V c_3^{p-6} e^{\phi(p/4-1/2)}+\Delta^{1/2}(p-7)(p-8)\Bigr] 
\bigg|_{t=0}=0\,, 
\eqlabel{conste} 
\end{equation} 
thus making \eqref{init} inconsistent. As a result, there are no smooth  
time-reversal invariant solutions realizing S-branes in the coupled tachyon  
matter-supergravity system with $\ISO(p+1)\times\SO(8-p,1)$ symmetry. 
 
\subsection{A  singularity theorem for $p<8$} 
 
Here we present a straightforward generalization of the singularity  
theorem of sections (2.3), (2.4) applicable for $p<8$. As before, we will  
show that the average Hubble parameter $H$ given by \eqref{hdef} 
will diverge in finite time, provided the initial condition  
\eqref{hbound} holds. 
 
Consider the general linear combination 
\begin{equation} 
\begin{split} 
k_1\times (\ref{FRW}.1)+ k_2 &\times (\ref{FRW}.2)+ k_3\times (\ref{FRW}.3)\\ 
&\Updownarrow\\ 
LHS&=RHS\,, 
\end{split} 
\eqlabel{comb} 
\end{equation} 
where $k_i$ are some (for now arbitrary!) constants and the additional  
index $i$ in $(\ref{FRW}.i)$ refers to the $i$-th  equation from  
the top in \eqref{FRW}. We want to rewrite the resulting equation \eqref{comb}  
so that it involves only derivatives of $H$ as defined in \eqref{hdef}.  
It is easy to see that the vanishing of  $\dot\Hpe$ (after eliminating  
$\dot\Hpp$) requires  
\begin{equation} 
k_3=\frac{8-p}{p+1} k_2\,. 
\eqlabel{k3} 
\end{equation} 
Now, the left hand side $LHS$ of \eqref{comb} 
can be combined as  
\begin{equation} 
\begin{split} 
LHS=&-9 \frac{k_1(p+1)-k_2}{p+1}\dot H-9 \frac{k_1(p+1)-9 k_2}{p+1}\ H^2  
\\ 
&-\frac{k_2(p-7)(p-8)}{\ape^2(p+1)}-\frac{k_1(p+1)(8-p)}{9}(\Hpp-\Hpe)^2 
\,. 
\end{split} 
\eqlabel{LHS} 
\end{equation} 
The right hand side of \eqref{comb} reads 
\begin{equation} 
\begin{split} 
RHS=&\frac{k_1(7-p)+k_2}{16}e^{a\phi} A^2+\frac {k_1}{2}\left( 
\dot\phi\right)^2\\ 
&+\frac{(k_1(p+1)-k_2)e^{(p-4)\phi/4}}{16 
\sqrt{\Delta}} V\lambda \ape^{p-8}\left(\dot T 
\right)^2\\ 
&+\frac{(k_1(p+1)(6-p)+k_2(10+p))e^{(p-2)\phi/4}}{16 
\sqrt{\Delta}(p+1)} V\lambda \ape^{p-8}\,. 
\end{split} 
\eqlabel{RHS} 
\end{equation} 
The equation \eqref{comb} takes the form  
\begin{equation} 
-9 \frac{k_1(p+1)-k_2}{p+1}\dot H=9 \frac{k_1(p+1)-9 k_2}{p+1}\ H^2+ 
[{\rm smth}]\,, 
\eqlabel{combr} 
\end{equation} 
where $[{\rm smth}]$ can be easily deduced from \eqref{LHS} and \eqref{RHS}. 
We will be interested in the following conditions on the $k_i$ 
\begin{align} 
&k_1>0\,, && k_2> 0 , \notag\\ 
&k_1(p+1)-k_2 >0, && k_1(p+1)-9 k_2\ge 0,\eqlabel{condit}\\ 
&k_1(7-p)+k_2 \ge 0, && k_1(p+1)(6-p)+k_2(10+p)\ge 0\,. 
\notag 
\end{align} 
Note that we can always find $k_1,k_2$ satisfying \eqref{condit} 
(notice that some conditions are $\ge$ and some other are $>$ --- 
this is important!) provided $p<8$.  
Now, given \eqref{condit}, $[{\rm smth}]$ in \eqref{combr} is always  
nonnegavite. So if the initial conditions for the  
universe are such that $H|_{t\to -\infty}<0$, we can replace  
\eqref{combr} with  
\begin{equation} 
-9 \frac{k_1(p+1)-k_2}{p+1}\dot H=9 \frac{k_1(p+1)-9 k_2}{p+1}\ H^2\,. 
\eqlabel{newcomb} 
\end{equation} 
Again, the point is that if $H$ satisfying \eqref{newcomb} will diverge  
in finite time, $H$ satisfying \eqref{combr} will diverge  
earlier. It is trivial to solve \eqref{newcomb} and see the divergence 
\begin{equation} 
H=\frac{k_1(p+1)-k_2}{t (k_1(p+1)-9 k_2)+\delta}\,, 
\eqlabel{solH} 
\end{equation} 
where $\delta$ is an integration constant.  
 
\subsection{Comparison with Leblond-Peet.} 
 
The main result of this note is in apparent contradiction with the recent  
claim of the construction of singularity-free S-brane solutions in  
supergravity for the maximally symmetric ansatz of \cite{BLW} made by  
Leblond and Peet in \cite{LP}. In this section we attempt to understand
this contradiction.

{} From the analytic side, we have been able to trace back the discrepancy 
to what we believe is an incorrect implementation of the smearing. As we 
have explained in the text, the only smearing consistent with the symmetries 
is uniformly in the transverse space, but constant in the parallel directions. 
In particular, $d(\r_0)/dt=0$, where $\r_0$ is as in \eqref{rour}. This condition 
is not properly implemented in \cite{LP}, as is apparent from their equation 
(3.30)\footnote{We are referring to the numbering of equations in 
hep-th/0303035v1.}. It appears that while some of the equations (3.31)-(3.42) 
do satisfy this condition, others do not, and this is one reason that their 
system of equations is inconsistent. For example, if $\rho_\perp$ is allowed 
to depend on the parallel directions, there should be a term containing 
$\del_\mu\r_{\perp}$ in the tachyon equation (3.23), and then a corresponding 
term later in (3.35). However, it is quite obvious that there is no way of making
$\rho_\perp$ depend on time in a way that renders the resulting equations 
consistent. A time dependent $\rho_\perp$ will simply not satisfy the energy  
conservation condition \eqref{consistency}. (One can also view 
this as conservation of the number of unstable branes.) To be fair, 
we wish to acknowledge that the corresponding equation in \cite{BLW}  
(eq.\ (10)), is somewhat imprecise. The precise meaning of this 
equation, which is also the meaning used in the equations of motion 
and singularity analysis of \cite{BLW}, is as in eqs.\ \eqref{rhop}  
and \eqref{rour} of the present note. 
 
Note that in \eqref{explicit}, the Maxwell equation can be explicitly 
integrated, to give 
\begin{equation} 
A = \frac{Q + \lambda F(T)}{\ee^{a\phi} c_3^{8-p}} \,, 
\eqlabel{integrate} 
\end{equation} 
where $F =\int f $, and $Q$ is an integration constant. Eq.\  
\eqref{integrate} has the desirable physical interpretation that the 
total ``charge of the S-brane'' (the asymptotic value of the RR-field) 
can be completely determined from the initial conditions on the tachyon. 
One of the consequences of the time-dependence of $\rho_\perp$ in 
\cite{LP} is that the Maxwell equation can {\it not} be integrated  
anymore in this way. 
 
For completeness, we now write the equations in string frame. The  
transformation to string frame in the notation of \cite{LP} is 
\begin{equation} 
\begin{split} 
&c_1=e^{-\phi/4},\qquad c_2=a e^{-\phi/4},\qquad c_3=R e^{-\phi/4},\qquad  
A=c_1^{-1} c_2^{-p-1}\ \dot{C}\,. 
\end{split} 
\eqlabel{trans} 
\end{equation} 
Eq.(3.31) of \cite{LP} becomes 
\begin{equation} 
{\ddot{C}}+ {\dot{C}}\left[(8-p){\frac{{\dot{R}}}{R}} 
-(p+1){\frac{{\dot{a}}}{a}}\right] = \lambda a^{p+1}R^{p-8} 
f(T){\dot{T}}. 
\end{equation} 
Eq.(3.34) of \cite{LP} becomes 
\begin{equation} 
\begin{split} 
& {\ddot{\Phi}} +{\dot{\Phi}} 
\left[(8-p){\frac{{\dot{R}}}{R}}+(p+1){\frac{{\dot{a}}}{a}} 
\right] -2{\dot{\Phi}}^{2} \nonumber\\ & = {\frac{(3-p)}{4}} 
\left({\frac{e^\Phi {\dot{C}}}{a^{(p+1)}}}\right)^2 
+{\frac{\lambda}{4}} e^{\Phi} V(T) R^{p-8}\left[(3-p)\sqrt{\Delta} - 
{\frac{1}{\sqrt{\Delta}}} \right] \, . 
\end{split} 
\end{equation} 
Eq.(3.35) of \cite{LP} becomes 
\begin{equation} 
{\ddot{T}} =  \Delta \left\{ {\dot{\Phi}}{\dot{T}} - 
{\dot{T}}\left[ (p+1){\frac{\dot{a}}{a}} \right] - 
{\frac{1}{V(T)}}{\frac{dV(T)}{dT}} - {\frac{f(T)}{V(T)}} {\dot{C}} 
e^{\Phi} a^{-(p+1)}\sqrt{\Delta} \right\} \,, 
\end{equation} 
Note the sign typo for $f/V$ term. Finally, the Einstein equations in 
\cite{LP} (3.40)-(3.42) are correct provided one makes the by now familiar  
replacement  
\begin{equation} 
\la\to \la R^{p-8}\,. 
\end{equation} 
We have explicitly verified that the equations of \cite{LP} modified as above  
are consistent. All conclusions we have reached in the main text about 
singularities in the geometry obviously hold here as well. 
 
Finally, let us comment that the numerical integration of the equations 
in \cite{LP} is rather problematic. Since the equations are inconsistent, 
one is bound to get different results depending on what subset of equations 
one chooses to numerically integrate. In any case, it also appears \cite{forth} 
that the code used in \cite{LP} is unstable and this appears to be another possible 
reason for the contradiction.
 
\end{appendix} 
 
\providecommand{\href}[2]{#2}\begingroup\raggedright 
\endgroup

\end{document}